\documentclass[aps,floatfix,superscriptaddress,prl,twocolumn]{revtex4}

\usepackage{amsmath, amsthm, amssymb, graphicx}
\usepackage{color}
\usepackage{multirow}
\usepackage{ulem}
\usepackage{float}
\usepackage{bm}
\usepackage{siunitx}
\usepackage{booktabs}
\usepackage{array}

\begin{document}
\title{The effect of the Magnus force on skyrmion relaxation dynamics}
\author{Barton L. Brown}
\affiliation{Department of Physics, Virginia Tech, Blacksburg, VA 24061-0435, USA}
\affiliation{Center for Soft Matter and Biological Physics, Virginia Tech, Blacksburg, VA 24061-0435, USA}
\author{Uwe C. T{\"a}uber}
\affiliation{Department of Physics, Virginia Tech, Blacksburg, VA 24061-0435, USA}
\affiliation{Center for Soft Matter and Biological Physics, Virginia Tech, Blacksburg, VA 24061-0435, USA}
\author{Michel Pleimling}
\affiliation{Department of Physics, Virginia Tech, Blacksburg, VA 24061-0435, USA}
\affiliation{Center for Soft Matter and Biological Physics, Virginia Tech, Blacksburg, VA 24061-0435, USA}
\affiliation{Academy of Integrated Science, Virginia Tech, Blacksburg, VA 24061-0405, USA}
\begin{abstract}
We perform systematic Langevin molecular dynamics simulations of interacting skyrmions in thin films.
The interplay between Magnus force, repulsive skyrmion-skyrmion interaction and thermal noise yields
different regimes during non-equilibrium relaxation. In the noise-dominated regime the Magnus force
enhances the disordering effects of the thermal noise. In the Magnus-force-dominated regime, the Magnus force
cooperates with the skyrmion-skyrmion interaction to yield a dynamic regime with slow decaying correlations.
These two regimes are characterized by different values of the aging exponent. In general, the Magnus force
accelerates the approach to the steady state.
\end{abstract}

\date{\today}

\maketitle

Magnetic skyrmions, particle-like spin textures encountered in many magnetic thin films and bulk materials with broken inversion symmetry 
and strong spin-orbit coupling under a weak applied magnetic field \cite{Rossler06,Muhlbauer09,Yu10,Yu11}, have recently been observed at 
room temperature \cite{Woo16,Boulle16}. This opens many possible avenues for applications in spintronics such as data storage \cite{Fert13,Kiselev11} 
and logic \cite{Zhang15} devices due to the ultra-low current densities required to move these topologically protected spin textures. 

Recent computational \cite{Lin13,Reichhardt16} and experimental \cite{Jiang17,Pollath17} evidence suggests that a particle-like treatment of skyrmions 
is valid in certain regimes. Using Thiele's approach \cite{Thiele73,Nagaosa13}, which treats skyrmions as rigid point-like particles, 
equations of motion can be derived \cite{Lin13} that lend themselves to in-depth numerical simulations of interacting skyrmion systems.
A similar treatment has been used to derive equations of motion for vortices in type-II superconductors \cite{Nelson93,Assi16}, 
and the resulting equations are in fact quite similar to those for skyrmions \cite{Reichhardt14}, except for the Magnus force which is 
usually negligible in vortex dynamics. The Magnus force acts normal to the drift velocity of the skyrmion and can therefore cause orbits or spiraling trajectories.
Recent progress has focused on the steady-state properties and dynamical phase transitions of driven skyrmions moving in an environment with random quenched disorder 
\cite{Reichhardt15,Reichhardt16,Diaz17} or on a substrate \cite{Reichhardt15a,Reichhardt15b,Reichhardt17}.

As applications become more widespread it will be important
to develop a more complete understanding of the relaxation dynamics 
of interacting skyrmions. Exploiting the particle equations of motion and the resulting coarse-grained framework,
we probe in the following relaxation processes of many interacting skyrmions far from equilibrium. Our main emphasis is to gain a better
understanding of how the interplay between the Magnus force, the repulsive skyrmion-skyrmion
interaction, and thermal fluctuations affect the non-equilibrium relaxation properties of skyrmion systems.

\begin{figure}
\includegraphics[width=1\columnwidth,clip=true]{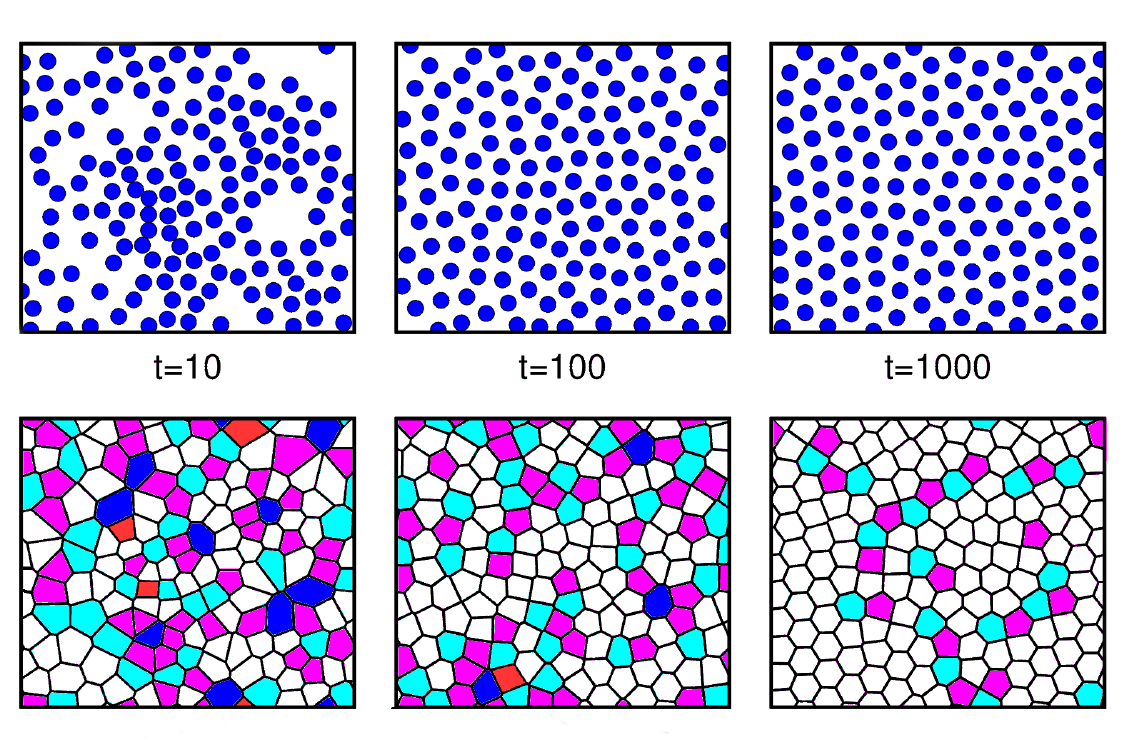}
\caption{A system of interacting skyrmions prepared in a disordered initial state at $t=0$ evolves over time into a triangular configuration.
Voronoi diagrams constructed from the particle positions are shown in the bottom row. In a perfect triangular lattice each region associated
with a particle would form a regular hexagon. At early times the Voronoi lattice is dominated by defects (polygons that are not hexagons), with the number 
of defects decreasing over time. Polygons have between four and eight sides and are colored in red, magenta, white, cyan, or blue from least to greatest side number.}
\label{Fig:system_state}
\end{figure}

In the absence of defects or strong noise, skyrmions crystallize into a triangular lattice configuration, see Fig. \ref{Fig:system_state}, 
due to the mutual repulsive force between them. Non-equilibrium relaxation kinetics of ordering systems are often paired with
physical aging phenomena. A many-body system is said to undergo physical aging if the following three properties are satisfied 
\cite{Henkel10}: (1) the relaxation towards equilibrium is slow, i.e. non-exponential; (2) time-translation invariance is broken; 
and (3) dynamical scaling is present. Prominent examples of physical aging can be
found in coarsening systems (including spin glasses and magnets), polymer glasses, and growth processes, to name but a few \cite{Henkel07,Rieger93,Henkel12}.
As revealed by numerous theoretical and experimental studies, physical aging is best studied through the investigation of two-time quantities \cite{Henkel10}.

We consider interacting skyrmions moving on a two-dimensional surface under the rigid structure approximation, 
where deformations of the internal spin structure are taken to be negligible 
so that the skyrmions are treated as point-like particles. Recent experiments point to the validity of this particle-like treatment \cite{Jiang17,Pollath17}.
The motion of $N$ interacting skyrmions is modeled according to a set of Langevin equations derived from Thiele's approach \cite{Thiele73,Nagaosa13,footnoteFP} and defined
in terms of the skyrmion drift velocities $\bm{v}_{i}(t)$ and positions $\bm{r}_{i}(t)$
\begin{equation}
\eta \bm{v}_{i} = \bm{F}^{M}_{i} + \bm{F}^{s}_{i} + \bm{f}~,
\label{Eq:particle_model}
\end{equation}
where $i=1,...,N$ labels the $N$ different skyrmions, whereas $\eta$ is the damping coefficient.
$\bm{F}^{M}_{i} = \beta \hat{z}\times \bm{v}_{i}$ is the Magnus force, whose strength can be adjusted by changing the value of the 
parameter $\beta$. The Magnus force, which acts in the direction perpendicular to the skyrmion's velocity, does not contribute
to the energy of the system and therefore does not break detailed balance.
Still, as shown in the following, the presence of the Magnus force significantly affects the transient 
dynamical properties of a system of interacting skyrmions.

The repulsive skyrmion-skyrmion interaction in (\ref{Eq:particle_model})
has the form (as determined through a numerical study \cite{Lin13})
$\bm{F}^{s}_{i}= \textstyle \sum_{i\neq j}F^{s}_{0}K_{1}(r_{ij}/\xi)\hat{r}_{ij}$ where $K_{1}(y)$ is the modified Bessel function of the second kind,
$\hat{r}_{ij} = \bm{r}_{ij}/\left| \bm{r}_{ij} \right| = (\bm{r}_{i} - \bm{r}_{j} )/\left| \bm{r}_{i} - \bm{r}_{j} \right|$ is the unit 
vector pointing from skyrmion $j$ to skyrmion $i$, and $\xi$ is the healing length. This force decays exponentially for $r_{ij}/\xi \gg 1$.
Finally, the last term on the right side is thermal white noise obeying $\langle f_{\mu}(t) \rangle = 0$ and 
$\langle f_{\mu}(t) f_{\nu}(t') \rangle = \sigma \delta_{\mu\nu}\delta(t-t')$, with $\sigma = 2\eta k_{B}T$ and $\mu$, $\nu = 1,2$
\cite{footnote}. No driving currents are
included here and the pinning due to defects are considered to be negligible. 

In the following simulations we take the skyrmion interaction coefficient $F^{s}_{0}=1$. We also apply a constraint on the system in terms 
of the coefficients $\eta$ and $\beta$, namely that $\eta^2+\beta^2=1$ \cite{Reichhardt15a,Reichhardt15b}. This constraint ensures that the 
average magnitude of the velocity of a free skyrmion is independent of the Magnus force. We also ran simulations without this
constraint (not shown) and verified that the same qualitative behavior is obtained as that discussed in the following. 
We choose units such that the healing
length $\xi = 1$ \cite{Lin13}.

We consider in the following systems of size $\frac{2}{\sqrt{3}} 36 \times 36$ with periodic boundary conditions that allow for the
skyrmions to form at equilibrium a triangular lattice. We tested that the skyrmions in the absence of noise indeed settle into this triangular configuration. 
The non-equilibrium simulations reported below have been done with $N=149$ skyrmions (which corresponds to a coverage of 10\%).
We checked that our results are robust against changing system sizes and/or coverage, as long as the particle picture remains valid.
As revealed by many other studies of non-equilibrium relaxation processes \cite{Henkel10}, results obtained for some system size are representative of
much larger systems as long as the system is not yet close to its steady state.
We assume that the system is initially in a disordered state where the skyrmions are located at random positions in our two-dimensional system.
The system is then allowed to
relax for $t>0$ at the temperature $T$ (or, equivalently, at a given value of $\sigma$) following the Langevin dynamics discussed above 
and quantities are measured as a function of time. 

\begin{figure}
\includegraphics[width=0.70\columnwidth,clip=true]{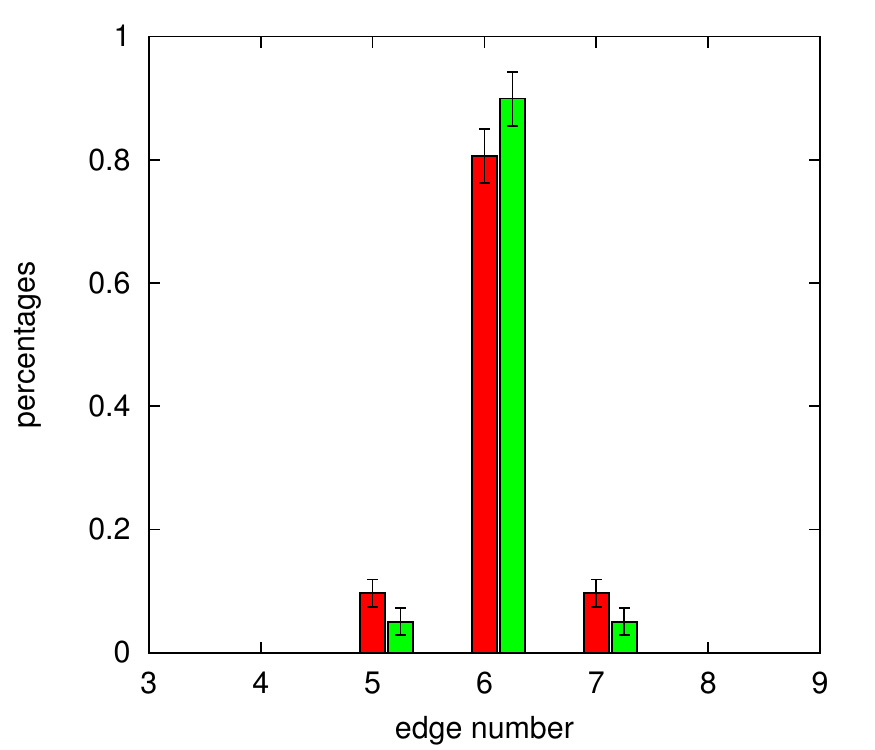}
\caption{Edge statistics of Voronoi diagrams generated from the skyrmion positions at time $t=800$ and noise strength $\sigma=0$, 
with the Magnus force either turned off (red/left) or on with $\beta / \eta = 5$ (green/right). An increase of the probability to observe
six-sided polygons is seen in simulations with a non-zero Magnus force,
indicating a faster relaxation into equilibrium in the presence of this force.
The data result from averaging aver 150,000 independent runs.}
\label{Fig:hist_comp}
\end{figure}

A first indication of the Magnus force's impact on relaxation processes can be garnered from a statistical analysis of Voronoi
maps like those shown in Fig. \ref{Fig:system_state}. In Fig. \ref{Fig:hist_comp} we compare for the case without noise the edge statistics 
for Voronoi diagrams obtained from simulations with (in green at the right)
and without (in red at the left) the Magnus force at time $t=800$ since preparation of the system. With the addition of the
Magnus force, six-sided domains become more likely than without, indicating that the presence of the Magnus
force facilitates the crystallization into the triangular lattice. This is a consequence of the fact that rearrangements of 
particles are easier due to the dynamical bending of the moving particles' trajectories. It is this facilitation of collective motion
that allows the system to find new, and quicker, paths towards the steady state.

For a more detailed investigation of the relaxation process and the related aging phenomena we turn to two-time quantities, such as the
two-time density autocorrelation function, that have been shown to provide valuable insights into these processes \cite{Henkel10}. 
Over time the repulsive skyrmion-skyrmion interaction tends to maximize the skyrmions' average 
nearest neighbor distance. As a result the skyrmions move away from their random initial positions, 
causing the decay of the density autocorrelation function.
In order to capture these changes we follow Ref. \cite{Pleimling11}
and set at time $s$ a circular area with radius $r$ (the data discussed
below have been obtained for $r=0.08$, but none of the observed features change when choosing a different, albeit similar, value for $r$) at the location of each skyrmion. 
At time $t>s$, we count the number of skyrmions still in their circles, generating the occupation numbers $n_{i}(t)$, with $n_{i}=0$ or $1$ (of course
$n_{i}(t=s)=1$ by construction). This quantity is then averaged over the $N$ different skyrmions as well as over many initial conditions
and realizations of the noise to produce the two-time autocorrelation 
\begin{equation}
C(t,s) = \Big \langle \frac{1}{N} \sum_{i=1}^{N}n_{i}(t)n_{i}(s) \Big \rangle~.
\label{Eq:den_ac}
\end{equation}

\begin{figure} \includegraphics[width=1\columnwidth,clip=]{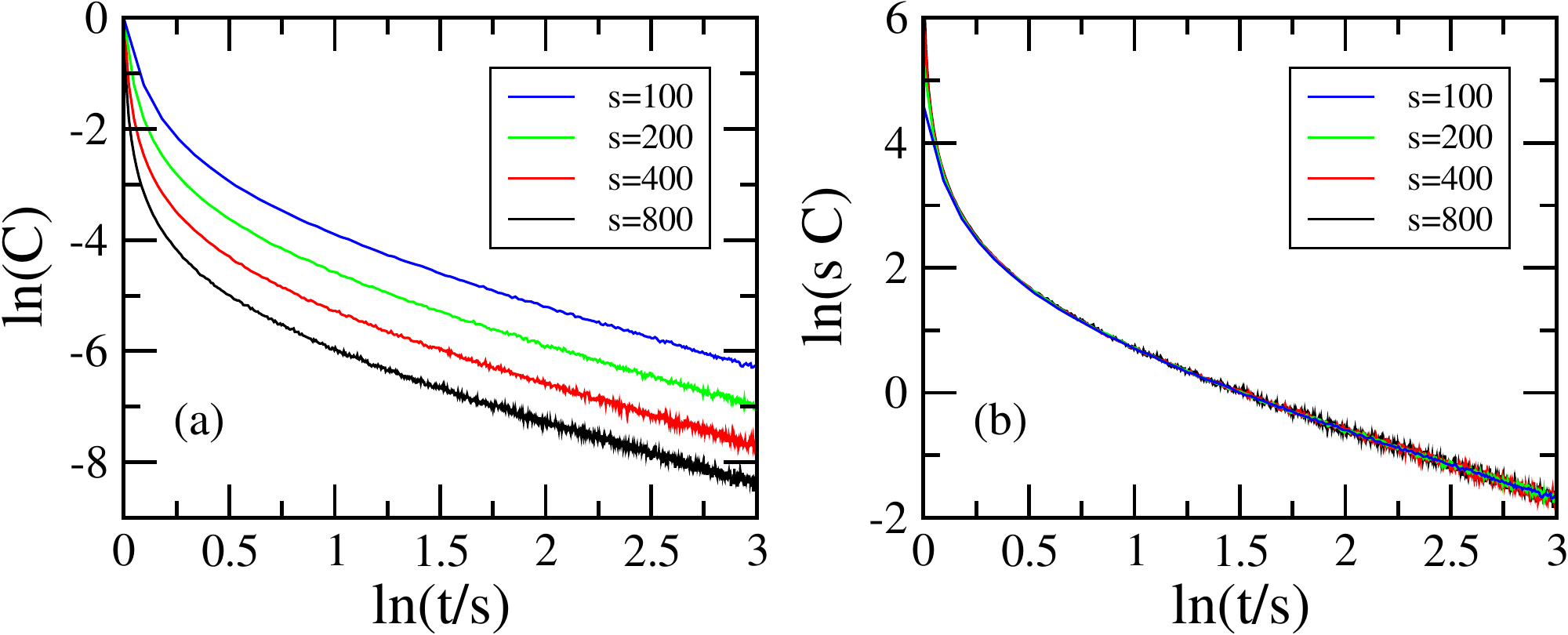}
\caption{(a) The two-time autocorrelation (\ref{Eq:den_ac}) of a system of non-interacting skyrmions, for various waiting times, $s$. 
The trajectories are completely determined by thermal fluctuations causing the skyrmions to perform random walks. 
This results in a simple aging scaling behavior with
an aging exponent $b=1$, see Eq. (\ref{Eq:scaling}), as shown in panel (b). The data displayed in this figure,
which result from averaging over 8000 independent runs, have been
obtained for $\sigma=0.1$ and $\beta / \eta = 0$, but the observed scaling is independent of the values of these
parameters.}
\label{Fig:free_skyrmion}
\end{figure}

A test case, which will help us to understand the relaxation properties of the
full problem, is provided by switching off the interactions between skyrmions, see Fig. \ref{Fig:free_skyrmion}.
In that case the skyrmions perform independent random walks, and it is expected that the two-time autocorrelation function (\ref{Eq:den_ac})
displays a simple aging scaling form \cite{Henkel10}
\begin{equation}
C(t,s)=s^{-b}f_C(t/s)~,
\label{Eq:scaling}
\end{equation}
with the aging exponent $b$ and a scaling function $f_C(y)$ that only depends on the time ratio $y=t/s$.
As shown in Fig. \ref{Fig:free_skyrmion}, the data obtained for different waiting times $s$
display a perfect scaling with $b=1$. We observe the same scaling for all values of $\sigma$, both in the absence or
presence of the Magnus force. Indeed the Magnus force, which corresponds to a non-zero value of $\beta$, serves only to rotate
the skyrmions in the plane but leaves the statistical properties of the random walks unaltered.

We supplement the results from the density autocorrelations by measuring also the time-dependent average nearest-neighbor distance
\begin{equation}
D(t) = \Big \langle \frac{1}{N}\sum_{i}^{N}\min_{i \neq j}r_{ij} \Big \rangle
\label{Eq:nearest_neighbor}
\end{equation}
from random initial conditions. In a regular triangular lattice where each of the $N$ sites is occupied by a skyrmion, the
distance between two neighboring skyrmions is $l_{\Delta}=\frac{2}{\sqrt{3N}}L$. For the system studied in this work
this yields $l_{\Delta}\approx 3.4$. The distance $D(t)$, which approaches the value $l_{\Delta}$ only at very late stages of the 
ordering process, provides a time-dependent length scale that may encode additional interesting details of the relaxation of the system
\cite{Pleimling15}.

\begin{figure}	\includegraphics[width=1\columnwidth,clip=true]{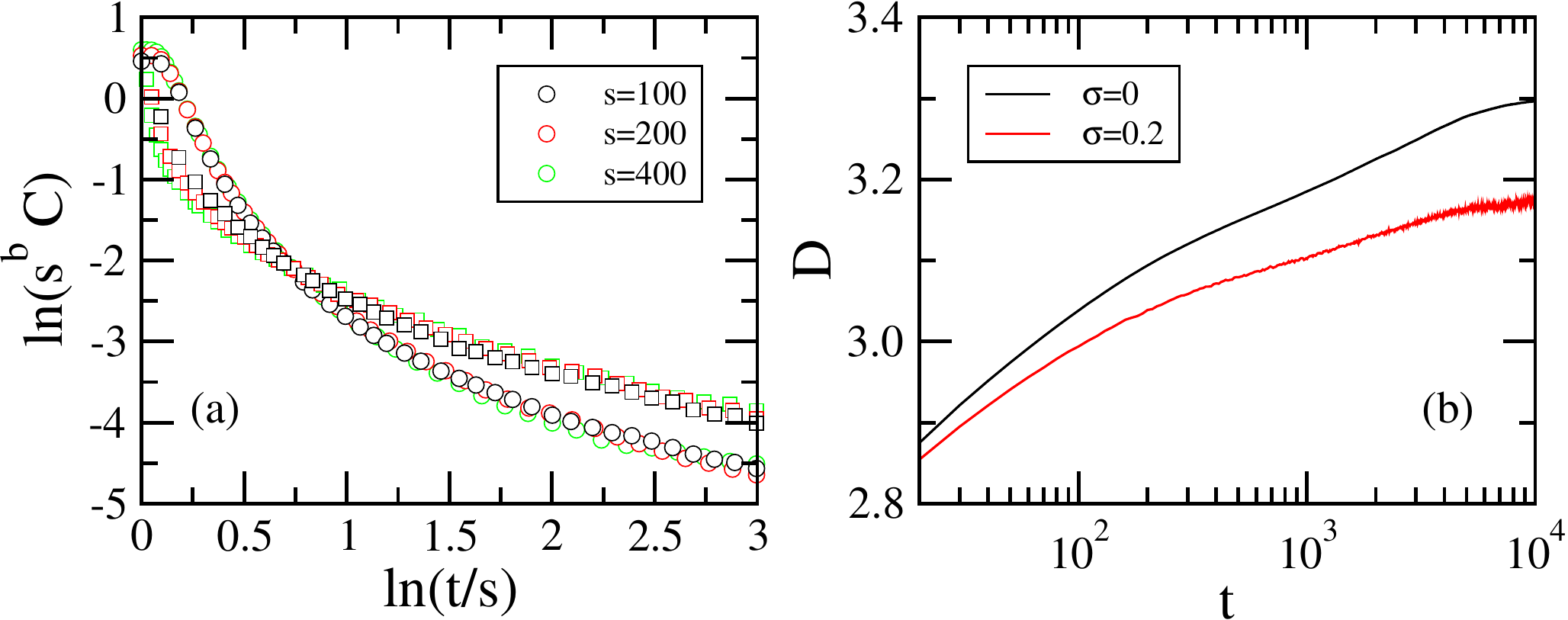}
\caption{(a) Scaled two-time autocorrelation function for different waiting times $s$ and (b) average nearest-neighbor distance for the two noise strengths $\sigma=0$ (circles in (a) and black line in (b)) and
$\sigma = 0.2$ (squares in (a) and red line in (b)) in the absence of the Magnus force. The scaling exponent $b$ in (a) varies with the noise strength $\sigma$,
with $b=0.10$ for $\sigma=0$ and $b=0.30$ for $\sigma = 0.2$.
At higher values of $\sigma$, skyrmions are on average driven further away from their preferred positions within the lattice 
which yields lower values of $D(t)$. For $\sigma = 0$ the system still contains imperfections at the end of our 
runs, with $D$ being slightly smaller than the equilibrium value. The data for the autocorrelation result from averaging over a
few thousands independent runs, whereas for $D(t)$ we averaged over a few hundred independent realizations.}
\label{Fig:M0_results}
\end{figure}

\begin{figure}
\includegraphics[width=1\columnwidth,clip=true]{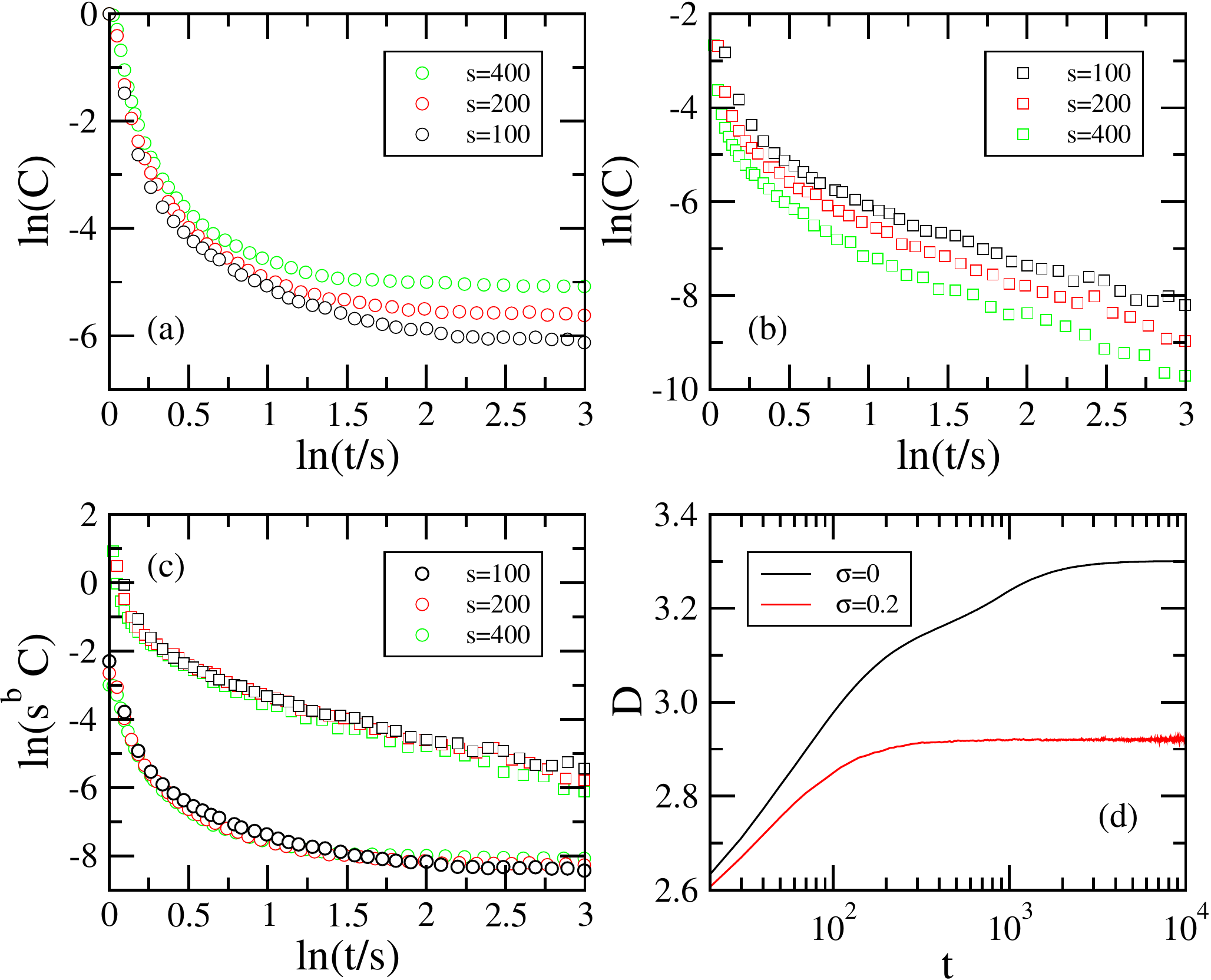}
\caption{Unscaled two-time autocorrelation function as a function of $t/s$ for different values of $s$
in the presence of a Magnus force with $\beta / \eta =5$: (a) $\sigma =0$ (circles) and (b) $\sigma = 0.2$ (squares).
In (a) the data sets are ordered from bottom to top when increasing the waiting time $s$, whereas in (b)
it is the other way round. (c) Scaled two-time autocorrelations ($\sigma =0$: circles; $\sigma = 0.2$: squares) and (d) average nearest neighbor distances 
for the same two cases. The value of the aging exponent changes sign when decreasing $\sigma$, being $b=0.60$ 
for $\sigma = 0.2$ and $b = - 0.50$ for $\sigma = 0$. 
}
\label{Fig:M05_results}
\end{figure}

The relaxation of interacting skyrmions is much more complex than that of free particles. As discussed in Figs.
\ref{Fig:M0_results} and \ref{Fig:M05_results}, the interplay between skyrmion interaction,
thermal fluctuations, and Magnus force yields aging scaling with non-universal (i.e. dependent on the strength
of the different terms) aging exponents. To disentangle the different contributions, we first discuss in Fig. \ref{Fig:M0_results} the
behavior in the presence of interactions and thermal noise alone, before considering in addition the effects of the Magnus force in Fig. \ref{Fig:M05_results}.

In the absence of pinning disorder our system of interacting skyrmions evolves toward the regular triangular lattice.
Thermal noise, however, prohibits the system to reach the perfect triangular lattice, which
instead settles for moderate noise levels into a partially ordered state where most of the particles fluctuate around their ground state positions.
Neglecting the Magnus force, this yields an average distance between skyrmions that is slightly reduced compared to that measured
in the absence of noise, see Fig. \ref{Fig:M0_results}b. Further increasing $\sigma$ increases this effect until for very large noise
levels (corresponding to large temperatures) the skyrmions behave effectively like free skyrmions as long as they are not too close.
Consequently, we expect for large values of $\sigma$ a scaling behavior close to that displayed in Fig. \ref{Fig:free_skyrmion}, with more
pronounced deviations showing up the smaller the value of $\sigma$ is. As shown in Fig. \ref{Fig:M0_results}a and Table \ref{Tab:scaling_exponent},
this is indeed the case, with the value of the aging exponent $b$ decreasing monotonically until it reaches the minimal value $b=0.10$
in the absence of noise.

\begin{table}
  \begin{tabular}{cc||cc}
    \toprule[1.5pt]
    \multicolumn{2}{c||}{Without Magnus Force} & \multicolumn{2}{c}{With Magnus Force}\\
    \multicolumn{2}{c||}{$\beta/\eta=0$} & \multicolumn{2}{c}{$\beta/\eta=5$}\\
    \hline
    $\sigma$ & $b$ & $\sigma$ & $b$\\
    \hline
   $\infty$ & 1.0 & $\infty$ & 1.0 \\
    0.5      & 0.45 & 0.5      &  0.95\\
    0.2      & 0.30 & 0.2      &  0.60\\
    0.1      & 0.20 & 0.1      & -0.20\\
    0.0      & 0.10 & 0.0      & -0.50\\
    \bottomrule[1.5pt]
  \end{tabular}
  \caption{Measured values of the aging scaling exponent with and without the Magnus force.}
  \label{Tab:scaling_exponent}
\end{table}

As already mentioned, a general property of the
Magnus force is to accelerate relaxation toward the final state. One way to see that is to compare Figs. 
\ref{Fig:M0_results}b and \ref{Fig:M05_results}d that show the time-dependent average distance between skyrmions.
In the presence of the Magnus force $D(t)$ approaches much faster a plateau-like behavior
that indicates the proximity to the final state. Focusing first on the case of large values of $\sigma$, 
we remark that in this noise-dominated regime the Magnus
force enhances the effects of the thermal noise, resulting in disordered states characterized by a much
smaller value of $D$ at long times than that encountered if the Magnus force is negligible (see the data for $\sigma = 0.2$). This is also readily
seen in the scaling of the autocorrelation, see Fig. \ref{Fig:M05_results}c and Table \ref{Tab:scaling_exponent}, where
the aging scaling exponent $b$ is much larger for $\sigma = 0.2$ and $\sigma = 0.5$ when the
Magnus force is present. Remarkably, for $\sigma = 0.5$ the value $b=0.95$ is already very close to the value $b=1$
for the case $\sigma = \infty$ where thermal noise completely dominates the skyrmion interaction.

A second regime of interest is that where the Magnus force dominates the thermal noise, yielding predominantly
curved trajectories. Comparing in Figs. \ref{Fig:M05_results}a and \ref{Fig:M05_results}b the
unscaled autocorrelation function for $\sigma = 0$ and $\sigma = 0.2$, we note an inversion of the order of the data sets
as a function of the waiting time. For $\sigma =0.2$, and this is the same for the cases without the Magnus force discussed
in Figs. \ref{Fig:free_skyrmion} and \ref{Fig:M0_results}, the larger the waiting time is, the smaller the value of $C$ is
for a given value of $t/s$, which yields a positive aging exponent $b > 0$. In contrast to this, for $\sigma = 0$ the autocorrelation
for a fixed value of $t/s$ is larger for larger waiting times, which then yields a {\it negative} aging exponent $b < 0$, see
Table \ref{Tab:scaling_exponent}. This remarkable change indicates that in the Magnus-force-dominated regime 
the complicity of the Magnus force and the skyrmion-skyrmion interaction yields states that are increasingly correlated
the closer the system gets to the steady state.

Our study of interacting skyrmions has yielded
important insights into relaxation processes for situations where the particle picture prevails.
These processes are heavily influenced by the interplay of the Magnus force, the repulsive skyrmion-skyrmion interactions,
and the thermal noise. In general, the Magnus force accelerates the relaxation process,
allowing the system to approach the steady state much faster than in the absence of this velocity-dependent force.
Our study reveals two different regimes, one dominated by thermal noise, the other dominated by the Magnus force.
In the former regime the Magnus force enhances the effects of the noise, resulting in final states with increased disorder,
as measured by the significantly lower value of the average distance between skyrmions. In the regime dominated by the
Magnus force, the cooperation between this force and the skyrmion-skyrmion interaction yields an increase of
the correlations between successive configurations. The change in regime is signaled by a change of the sign of
the aging exponent $b$.
It will be interesting to see whether these different regimes can be identified experimentally
in studies of interacting skyrmions at different temperatures.

This research was supported by the US Department of
Energy, Office of Basic Energy Sciences, Division of 
Materials Sciences and Engineering under Grant No. DE-FG02-09ER46613.



\begin{thebibliography}{99}
\bibitem{footnoteFP} Such a strongly interacting many-body system cannot be adequately captured by a Fokker--Planck equation.
\bibitem{footnote} As discussed in Ref. [10], values of the dimensionless parameters in
the Langevin equation (\ref{Eq:particle_model}) can be derived from experimental data. For example, 
for MnSi a temperature of 15 K corresponds to the value $\sigma=0.1$.
\end{thebibliography}
\end{document}